\documentclass[
 aip, 
reprint,
]{revtex4-1}

\usepackage[utf8]{inputenc}
\usepackage[T1]{fontenc}
\usepackage{diagbox}
\usepackage{amsfonts}
\usepackage{amsmath}
\usepackage{amsthm}
\usepackage{bm}
\usepackage{indentfirst}
\usepackage{microtype}
\usepackage[squaren]{SIunits}
\usepackage{booktabs}
\usepackage{amssymb}
\usepackage{graphicx}
\usepackage{multirow}
\usepackage{wallpaper}
\usepackage[normalem]{ulem}
\newcommand{\bra}[1]{\langle #1|}
\newcommand{\ket}[1]{|#1\rangle}

\newcommand{\beq}{\begin{eqnarray}}
\newcommand{\eeq}{\end{eqnarray}}
\newcommand{\bea}{\begin{eqnarray}}
\newcommand{\eea}{\end{eqnarray}}
\newcommand \avec{{\bf a}}
\newcommand \kvec{{\bf k}}

\newcommand \qvec{{\bf q}}

\newcommand\rvec{{\bf r}}

\newcommand\Rvec{{\bf R}}

\newcommand\Avec{{\bf A}}

\usepackage{braket}
\usepackage{caption}
\usepackage{mathtools}
\usepackage{subcaption}
\usepackage[colorlinks=true,urlcolor=blue,linkcolor=blue,
            citecolor=blue]{hyperref}
\usepackage[normalem]{ulem}
\newcommand\redsout{\bgroup\markoverwith{\textcolor{red}{\rule[.5ex]{2pt}{0.4pt}}}\ULon}
\newcommand{\bb}[1]{{\mathbf{#1}}}

\def\simge{\mathrel{%
       \rlap{\raise 0.511ex \hbox{$>$}}{\lower 0.511ex \hbox{$\sim$}}}}
\def\simle{\mathrel{
       \rlap{\raise 0.511ex \hbox{$<$}}{\lower 0.511ex \hbox{$\sim$}}}}

\begin{document}


\title{Electronic structure and optical properties of quantum crystals from 
first principles calculations in the Born-Oppenheimer approximation}

\author{Vitaly Gorelov}
\affiliation{Maison de la Simulation, CEA, CNRS, Univ. Paris-Sud, UVSQ, Universit{\'e} Paris-Saclay, 91191 Gif-sur-Yvette, France}
\author{David M. Ceperley}
\affiliation{Department of Physics, University of Illinois, Urbana, Illinois 61801, USA}
\author{Markus Holzmann} 
\affiliation{Univ. Grenoble Alpes, CNRS, LPMMC, 3800 Grenoble, France}
\affiliation{Institut Laue Langevin, BP 156, F-38042 Grenoble Cedex 9, France}
\author{Carlo Pierleoni}
\affiliation{Maison de la Simulation, CEA, CNRS, Univ. Paris-Sud, UVSQ, Universit{\'e} Paris-Saclay, 91191 Gif-sur-Yvette, France}
\affiliation{Department of Physical and Chemical Sciences, University of L'Aquila, Via Vetoio 10, I-67010 L'Aquila, Italy}
\date{\today}

\begin{abstract}
We develop a formalism to accurately account for the renormalization of electronic structure due to quantum and thermal nuclear motions within the Born-Oppenheimer approximation. 
We focus on the fundamental energy gap obtained from electronic addition and removal energies from Quantum Monte Carlo calculations in either the canonical or grand canonical ensembles. 
The formalism applies as well to effective single electron theories such as those based on Density Functional Theory.
We show that electronic (Bloch) crystal momentum can be restored by
marginalizing the total electron-ion wave function with respect to the nuclear equilibrium distribution, 
and we describe an explicit procedure to establish the band structure of electronic excitations for quantum crystals within the Born-Oppenheimer approximation. 
Based on the Kubo-Greenwood equation, we discuss the effects of nuclear motion on optical conductivity. 
Our methodology applies to the low temperature regime where nuclear motion is quantized and in general differs from the semi-classical approximation.
We apply our method to study the electronic structure of C2/c-24 crystalline hydrogen at 200K and 250 GPa 
and discuss the optical absorption profile of hydrogen crystal at 200K and carbon diamond at 297K. 
\end{abstract}

\maketitle

\section{\label{sec:Intro}Introduction}

With increasing computational power, precise Quantum Monte Carlo (QMC) calculations of electronic properties in real materials have become affordable \cite{PhysRevLett.105.086403,Koloren__2011,PhysRevB.88.245117,PhysRevB.90.125129,Wagner_2016,PhysRevB.98.075122,Yang2020,PhysRevB.101.165124,PhysRevB.101.165125}. 
Quantum Monte Carlo methods naturally extend to solve the full Schr{\"o}dinger equation of the coupled
electron-ion system at zero temperature \cite{PhysRevB.36.2092,PhysRevLett.70.1952,CPC}
or at finite temperature within the Path Integral formalism \cite{PhysRevLett.73.2145,Magro1996,PhysRevLett.85.1890}. 
For typical temperatures in condensed matter, the Born-Oppenheimer approximation can be further used to sample the nuclear distribution either
within coupled electron ion Monte Carlo (CEIMC) \cite{PhysRevLett.93.146402} or molecular dynamics simulations \cite{PhysRevLett.94.056403,PhysRevLett.100.114501}. 

Recently, we studied the effect of nuclear quantum and thermal motion on the closure of the fundamental electronic gap in high-pressure solid and liquid hydrogen within CEIMC \cite{Gorelov2020,gorelov2020electronic}. 
Here, we discuss how the fundamental gap and the band structure can be obtained for quantum and thermal crystals in a fully non-perturbative approach using QMC based methods. 
We further propose a new scheme to effectively include nuclear quantum and thermal effects on optical properties at low temperature where the usual semiclassical approximation breaks down. 

Standard electronic structure methods based on effective single electron theories like Density Functional Theory (DFT), or many-body perturbation theory (GW), 
often assume a weak electron-phonon coupling and phonons within the harmonic approximation so that their effects on the electronic structure can be treated perturbatively \cite{Giustino2010,Marini2008,Cannuccia2011,Cannuccia2012,Antonius2014,Ponce2014,Kawai2014,Ponce2015,MolinaSanchez2016,Menendez2017,QueralesFlores2019,Lihm2020,Zacharias2010_2}, 
based on the seminal work of Allen, Heine, and Cordona \cite{Allen1976,Allen1981,Allen1983}.
Both assumptions \cite{Giustino2017} limit the predictive power of these methods, especially for systems of light nuclei, 
like solid hydrogen \cite{PhysRevB.87.184107,natcom} and other molecular crystals \cite{Monserrat2015}. 
To go beyond the harmonic limit, one can employ the self-consistent harmonic approximation (SCHA), 
an effective mean field theory based on minimizing a free energy bound with respect to the effective equilibrium nuclear positions 
and effective frequencies within an harmonic ansatz for the nuclear motion. 
One characteristic of SCHA is to allow for structural predictions induced by nuclear quantum and thermal effects
\cite{Errea2014,Monacelli2020}, but is intrinsically difficult to improve on further.

Non-perturbative treatment of phonons can be achieved by path-integral calculations of the nuclear motion within the Born-Oppenheimer approximation. 
Path integral molecular dynamics (PIMD) and Monte Carlo (PIMC) has been used to 
study the renormalization of electronic structure of different materials 
due to nuclear motion mainly based on a semiclassical interpretation of the instantaneous Born-Oppenheimer electronic energies of the nuclear trajectory \cite{DellaSala2004,Ramirez2006,Franceschetti2007,Ramirez2008,Rillo2019,Morales2010,Pierleoni2016,Zacharias2020}. 
In this scheme, nuclear coordinates have been implicitly assumed to be good quantum numbers for the nuclear motion. 
However, this is questionable at low temperatures where quantization of phonons is important. 
Here, we will show how to take properly into account the renormalization of electronic structure at low temperatures within PIMD and PIMC extending the discussion anticipated in Ref.~\cite{Gorelov2020}.

Concerning optical properties, \textit{ab--initio} calculations based on the Born-Oppenheimer approximation typically employ the semi-classical William-Lax approach \cite{Williams1951,Lax1952}, 
where the optical spectra computed at fixed nuclear configuration are averaged over the equilibrium nuclear distribution. 
Successfully applied to study optical properties of heavier elements \cite{Patrick2014,Zacharias2016}, its
accuracy for light elements like hydrogen remains questionable and we will discuss
the minimal changes needed at low temperatures when nuclear motion is quantized.

The paper is organized as follows. 
In section \ref{sec:Gap} we discuss the energy gaps and band structure of quantum crystals in a many-body framework. 
We first define the fundamental gap based on electron addition and removal energies within the BO approximation in the canonical and grand canonical ensemble.
We then show how the concept of crystal momentum of electronic excitations can be
meaningfully extended to the case of quantum crystals in a fully non-perturbative way, and finally present some illustrative results for solid hydrogen at 200 K and 250 GPa. 
In section \ref{sec:OptProp} we then study the effects of finite temperature and zero point nuclear motion on the optical properties for semiclassical and quantum nuclei 
and compute optical absorption of solid hydrogen and carbon diamond at finite temperature. In section \ref{sec:Concl} we summarize our results. 
Explicit steps involved in the calculations of the density of states (DOS) in the grand canonical ensemble are detailed in the Appendix.

\section{\label{sec:Gap}Energy gap and band structure}

In the following we define and discuss single particle electronic excitation
energies focusing on the fundamental electronic energy gap and the band structure
in a quantum crystal at zero and finite temperatures. We will assume 
the validity of the Born-Oppenheimer approximation, a simplification usually well justified in the description of condensed-matter. 
We stress that the formalism equally applies to many-body and effective single-electron theories.

\subsection{\label{sec:GapCan}Canonical ensemble}

Let us consider a system with $N_p$ protons and $N_e$ electrons, mutually interacting via the bare Coulomb interaction. To assure global charge neutrality, a uniform charge background
is added whenever $N_p \ne N_e$. 
The canonical partition function at nuclear temperature $T=1/k_B \beta$  and
volume $V$ can be written as a 
path-integral
\beq
Z(N_e)=e^{- \beta F(N_e)}= \int {\cal D}\bb{R}(\tau) e^{-S[\bb{R}(\tau)]}
\eeq
where the action 
is 
\beq
S[\bb{R}(\tau)]=\int_0^\beta d\tau \left[ \frac{\hbar^2}{2M} 
\left(\frac{d\bb{R}(\tau)}{d\tau} \right)^2
 + E_0(\bb{R}(\tau),N_e) \right].
\nonumber \\
\eeq
Here $M$ is the nuclear mass, $\bb{R}$ the vector of the $N_p$ nuclear coordinates,
and $E_0(\bb{R},N_e)$ denotes the electronic Born-Oppenheimer 
potential energy surface (PES) calculated with $N_e$ electrons. As we will concentrate on situations where
we only vary the number of electrons $N_e$ keeping $V$, $T$, and
$N_P$ constant, we do not explicitly write out the dependence on these latter variables.

Within CEIMC calculations the imaginary time discretized path-integral is typically calculated
for the undoped situation, $N_e=N_p$, and the free energy difference can be obtained from
\beq
\frac{Z(N_e)}{Z(N_p)}= e^{- \beta [ F(N_e)-F(N_p)]} 
= \left\langle e^{- \int_0^\beta d\tau \delta E_0(\bb{R}(\tau),N_e) }\right\rangle
\nonumber \\
\eeq
where $\delta E_0(\bb{R},N_e)=E_0(\bb{R},N_e)-E_0(\bb{R},N_p)$
and $\langle \cdots \rangle$ denotes the averaging over the undoped sample, $N_e=N_p$.

\begin{figure}[t]
\center
\begin{minipage}[b]{\columnwidth}
\includegraphics[width=\columnwidth]{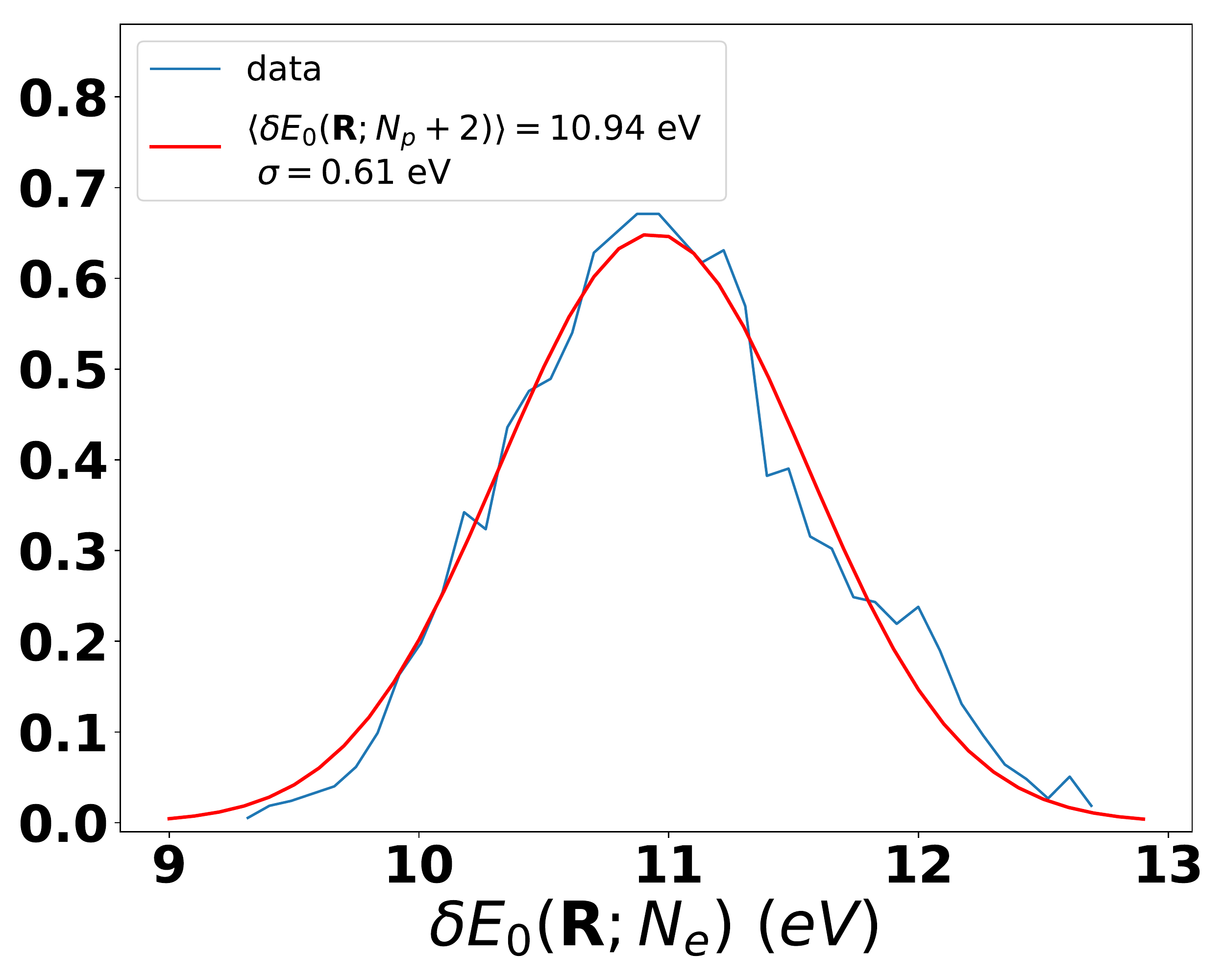}
\end{minipage}
\caption{\small{
Histogram of the distribution of energy needed to add two electrons to  C2/c-24 solid hydrogen at 200 K and 248 GPa 
compared with a normal distribution.}
\label{fig:NormCheck}}
\end{figure}

For small doping, $|N_e-N_p| \ll N_p$, 
the potential energy surface should only be
slightly perturbed,
 $|\delta E_0(\bb{R},N_e)| \ll |E_0(\bb{R},N_p)|$, and
we may use the cumulant expansion (see Fig.~\ref{fig:NormCheck}),  which gives
\beq
F(N_e)-F(N_p) = 
\langle \delta E_0(\bb{R}(0),N_e) \rangle -\frac{\sigma^2(N_e)}{2} 
\label{df}
 \\
\sigma^2(N_e)= \int_0^\beta d \tau \langle \delta E_0(\bb{R}(\tau),N_e) \delta E_0(\bb{R}(0),N_e)
\rangle_c 
\eeq
where $\langle\dots\rangle_c$ indicates the centered cumulant.
Having the form of a static response function, $\sigma^2(N_e)$ describes the leading order
changes of the adding/removal energies due to the electronically doped potential energy surface.

Adding (or removing) a single electron, we obtain the chemical potential from the free energy
differences
\beq
\mu_{\pm }= F(N_p\pm 1)-F(N_p).
\eeq
A gap with respect to electronic doping,
$\Delta=\mu_+-\mu_-$, can be determined from thermodynamics.

Time-correlations of electron addition/removal energies described by
 $\sigma^2(N_p\pm 1)$ contribute with opposite sign to the gap
and are expected to largely cancel each other, 
\beq
\Delta \simeq \langle \delta E_0(N_p+1)\rangle - \langle \delta E_0(N_p-1)\rangle
\eeq
so that the gap is entirely determined by the electronic addition and removal energies, as intuitively expected. 

\subsection{\label{sec:GapGrandCan}Grand canonical ensemble}

In the grand canonical ensemble,  instead of the electronic density, the (electronic) chemical
potential $\mu$ is the independent variable. Since the integrated density of states is given by the density as a function $\mu$,
a non-vanishing gap implies a vanishing of the 
density of states in the region $\mu_- < \mu < \mu_+$.

The use of twist averaged boundary conditions provides an efficient strategy to
reduce the finite size effects that result from the discrete nature of filling single particle orbitals \cite{Lin2001,Chiesa2006,PhysRevB.78.125106,Holzmann2016,PhysRevB.100.245142}.
In the following we describe its use in the grand canonical approach of calculating
the fundamental gap \cite{Yang2020,Gorelov2020}.

Expressed in terms of the canonical partition function $Z(N_e,\theta)$,
the grand canonical partition function  is
\beq
Z_{gc}(\mu,\theta) &= &\sum_{N_e} e^{\beta \mu N_e} Z(N_e,\theta)
\nonumber \\
&=& \sum_{N_e} e^{-\beta ( F(N_e,\theta)-\mu N_e)}
\label{zgc}
\eeq
where we have explicitly added the dependence on the twist angle $\theta$
imposed on the boundary conditions of the electronic wave functions.

Consistent with the BO approximation, we address situations where the temperature
is much less than the electronic excitation energy, so that in equilibrium one maximises the exponent in
Eq.~(\ref{zgc})
\beq
-T\log Z_{gc}(\mu,\theta) &= & \text{min}_{N_e} \left[F(N_e,\theta)-\mu N_e \right] 
\label{logZ} \\
&\equiv & F(N_e^\theta,\theta)- \mu N_e^\theta
\eeq
where $N_e^\theta$ denotes the number of electrons which minimizes the r.h.s. of Eq.~(\ref{logZ}).
Although not explicitly indicated, let us stress that the
finite temperature effects of the nuclear motion are  still contained in the temperature dependence of $F(N_e^\theta,\theta)$.

Neglecting modifications of the potential energy surface due to doping, e.g. $\sigma^2$
in the free energy differences Eq.~(\ref{df}) as discussed previously, $N_e^\theta$ can be determined by
replacing the free energy $F(N_e,\theta)$ with the electronic energy $E(N_e,\theta)$ in
the minimization Eq.~(\ref{logZ}).

To reduce finite size effects, different independent calculations are performed over a dense grid
of $M$ twist angles and the results are then averaged, 
\beq
f(\mu) &= &\frac{1}{M V} \sum_\theta F(N_e^\theta,\theta) \\
n_e(\mu) &= &\frac{1}{M V} \sum_\theta N_e^\theta  \label{eq:n_e} \\
e(\mu)& =&\frac{1}{M V} \sum_\theta E_0(N_e^\theta,\theta)
\eeq
In the independent particle approximation, the free energy, the internal energy and electronic densities, $f$, $e$, and $n_e$ respectively,
exactly agree with those obtained in a supercell of volume $MV$. In a many body theory, however, size effects
due to correlations of two or more electrons must be addressed differently \cite{Chiesa2006,Holzmann2016,Yang2020}.

A gap in the many-particle density of states for $\mu_-< \mu < \mu_+$ then implies 
a vanishing slope in fundamental thermodynamic properties
($f(\mu), n_e(\mu), e(\mu)$) as a function of $\mu$ in this region.
The value of the gap
\beq
\Delta=\mu_+ - \mu_-
\eeq
can be directly read off from the flat region in the plot of these functions.

We stress that the calculation of the thermodynamic potentials involves the average over nuclear configurations. This average should always be done before determining the fundamental gap.
Hence the correct value of the fundamental gap is, in general, different (larger) than the result 
obtained by determining the gap from the minimum value of electronic excitation energies with respect to all nuclear configurations in a simulation run. 
The result of the latter procedure, which we call the ``semiclassical approximation'', 
may be justified when analysing spectroscopic quantities 
but does not necessarily represent the thermodynamic gap.

\subsection{Electronic band structure in quantum crystals}\label{sec:SCtherm}

Within effective single particle theories, the electronic structure
of perfect crystals is conveniently described
by discussing the resulting band structure of the electronic excitation energies
as a function of their Bloch wave vector.
Since the Bloch wave vector characterizes the symmetry properties of the underlying nuclear
crystal, electronic addition and removal energies of a perfect crystal can still be labelled by its quasi-momentum also within a many-electron description \cite{PhysRevB.51.10591,PhysRevB.62.2330}. 

In the following we discuss the extension of the concept of quasi-momentum to the case of quantum
crystals, where the nuclear positions fluctuate around their perfect crystal lattice sites, $\Rvec_0$,
with zero point quantum and/or thermal nuclear motion.

\subsubsection{Harmonic approximation}
Let us first consider the case of small nuclear fluctuations around the lattice sites, such
that the harmonic approximation can be used. Treating the electron-ion interaction as a weak perturbation,
the $\alpha^{th}$ electronic wave function $\Phi_\alpha(\rvec|\Rvec)$ for a given (fixed) 
nuclear configuration $\Rvec$ is given in first order in perturbation theory as
\beq
\Phi_a(\rvec|\Rvec)\simeq \Phi_{\kvec 0}^{\Rvec_0}(\rvec) + \sum_{\qvec m} {}^{'} \frac{(\Rvec-\Rvec_0) \cdot 
\Avec_{\kvec 0}^{\qvec m}}{E_{\kvec 0}^{\Rvec_0}-E_{\qvec m}^{\Rvec_0}}
\Phi_{\qvec m}^{\Rvec_0}(\rvec).
\label{Pert}
\eeq
Here, $\Phi_{\kvec n}^{\Rvec_0}(\rvec) \equiv \langle \rvec |\Phi_{\kvec n}^{\Rvec_0} \rangle$ denotes the electronic (ground state) 
wave function
the perfect crystal, characterized by a (total) quasi-momentum $\kvec$ and a band index $n=0$. $\rvec$ represents all electronic coordinates. 
With nuclear displacements, 
$\Phi(\rvec|\Rvec)$ will have contributions from
electronic excited states $(\qvec m) \ne (\kvec 0)$ of the periodic Hamiltonian, and 
$\Avec_{\kvec 0}^{\qvec m} = 
\nabla_\Rvec \langle \Phi_{\qvec m}^{\Rvec_0} |H | \Phi_{\kvec 0}^{\Rvec_0} \rangle$
denotes the matrix elements evaluated at $\Rvec=\Rvec_0$. Only the electron-nuclear
interaction in the Hamiltonian $H$ contribute to this matrix elements.

Integrating Eq~(\ref{Pert}) over the distribution nuclear fluctuations with $\langle R-R_0\rangle=0$, e.g. zero-point or thermal fluctuations, the contribution of electronic excited states drops out:
\beq
\langle \Phi_\alpha(\rvec|\Rvec) \rangle = \phi_{\kvec 0}^{\Rvec_0}(r).
\eeq
The symmetry of the perfect crystal is thus restored, and, at least
in the harmonic approximation, we can characterize the $\alpha^{th}$
electronic wave function of the quantum crystal by the quasi-momentum
of the adiabatically connected state $\kvec$,
\beq
\langle \Phi_\alpha(\rvec+\avec|\Rvec) \rangle = e^{i \kvec \cdot \avec}
\langle \Phi_\alpha(\rvec|\Rvec) \rangle
\eeq
where $\avec$ denotes any of the crystal lattice vectors.

In general, the electronic ground state of the system will have a vanishing
quasi-momentum in a large enough supercell, but
adding and removal energies are still characterized by
a quasi-momentum within the first Brillouin zone (BZ) even for a quantum crystal.
It remains a useful concept for analysing transition matrix elements of
purely electronic operators: as long as the nuclear distribution remains
unchanged in the transition matrix, the quasi-momentum must be conserved, and
the usual selection rules for perfect crystals apply. In particular, one
may distinguish between direct and indirect transition according to the
Bloch vectors involved.

\subsubsection{Beyond the harmonic approximation}

In general, the crystal lattice periodicity implies a discrete translation symmetry
such that a combined translation of electron and nuclei
by any lattice vector does not change the
many-body density of system. Therefore, any eigenfunction of the full many-body system,
$\Psi_{\kvec n}(\rvec,\Rvec)$,
 can be characterized
by a Bloch vector $\kvec$ in the first BZ
\beq
\Psi_{\kvec n}(\rvec+\avec,\Rvec+\avec) = e^{i \kvec \cdot \avec} \Psi_{\kvec n}(\rvec,\Rvec)
\eeq
The quasi-momentum transfers to any electronic wave function obtained by
marginalizing the full wave function with an arbitrary nuclear wave function 
$\chi(\Rvec+\avec)=\chi(\Rvec)$, 
\beq
\int d \Rvec \chi^*(\Rvec) \Psi_{\kvec n}(\rvec+\avec,\Rvec)  
= e^{i \kvec \cdot \avec} \int d \Rvec \chi^*(\Rvec) \Psi_{\kvec n}(\rvec,\Rvec).
\nonumber \\
\eeq
This allows us to extend the concept of the quasi-momentum of electronic excitations and
to establish a band structure
useful for discussing the character of transition matrix
elements for quantum crystals beyond the harmonic approximation.

\subsubsection{Born-Oppenheimer approximation}

Determining the quasi-momentum of electronic wave functions in the Born-Oppenheimer approximation
based on the above considerations
is not straightforward. Within the Born-Oppenheimer approximation, the electronic wave function, $\Phi_\alpha^{\Rvec}(\rvec)=\langle \rvec| \Phi_\alpha^\Rvec\rangle$,
is only determined up to an arbitrary phase $\varphi(\Rvec)$, which depends on the 
nuclear positions
\beq
\Phi_\alpha^\Rvec(\rvec) \simeq \Phi_\alpha(\rvec|\Rvec) e^{i \varphi(\Rvec)}.
\label{BOphase}
\eeq
Although $\varphi(\Rvec)$ can be fixed (up to a global constant)
by requiring smooth changes with respect to
adiabatic changes of the nuclear position,
this is impractical for numerical computations.

Let us therefore consider that the Born-Oppenheimer wave function, Eq.~(\ref{BOphase}),
up to the unknown phase $\varphi(\Rvec)$
exactly coincides with the harmonic expansion, Eq.~(\ref{Pert}) and expand it into
the components of the electronic eigenstates of the perfect crystal
\beq
\langle \Phi^{\Rvec_0}_{\qvec m} | \Phi_\alpha^\Rvec \rangle
& \simeq & \left[ \delta_{\qvec \kvec}\delta_{m0}
 + (1-\delta_{\qvec \kvec} \delta_{m0})
 \frac{(\Rvec-\Rvec_0) \cdot 
\Avec_{\kvec 0}^{\qvec m}}{E_{\kvec 0}^{\Rvec_0}-E_{\qvec m}^{\Rvec_0}}
\right]
\nonumber \\
&& \times e^{i \varphi(\Rvec) }
\label{coeff}
\eeq
The unknown phase $\varphi(\Rvec)$ can be eliminated considering the matrix elements
\bea
\bb{T}(\bb{q},m;\bb{q},m') = &&\nonumber \\
 \int d\Rvec |\chi(\Rvec)|^2
\left(\bb{R}-\bb{R}_{0}\right) 
  \langle \Phi_\alpha^\Rvec |\Phi^{\bb{R}_0}_{\bb{q}m'}\rangle 
\langle\Phi^{\bb{R}_0}_{\bb{q}m}| \Phi_\alpha^\Rvec \rangle
\nonumber \\ \label{eq:Tknqm}
=  \delta_{\bb{qk}}\Bigg[\delta_{m'0} \left(1-\delta_{m0}\right) \bb{t}_{\bb{k}0}^{\bb{q}m}
 + \delta_{m0} \left(1-\delta_{m'0}\right)  
\bb{t}_{\bb{k}0}^{\bb{q}m'} \Bigg] \nonumber \\
\eea
for $m \ne m'$ and
 $\bb{t}_{\bb{k}0}^{\bb{q}m}$ denotes a vector matrix element which can be identified with
the coefficients in the harmonic expansion on the r.h.s. of Eq.(\ref{coeff}). Again, any positive density distribution
$\chi(\Rvec)^2$ with $\langle \Rvec - \Rvec_0 \rangle=0$ for the nuclei can be used; in practice one simply averages over their
equilibrium distribution at zero or finite temperature.

\begin{figure}[t]
\center
\begin{minipage}[b]{\columnwidth}
\includegraphics[width=\columnwidth]{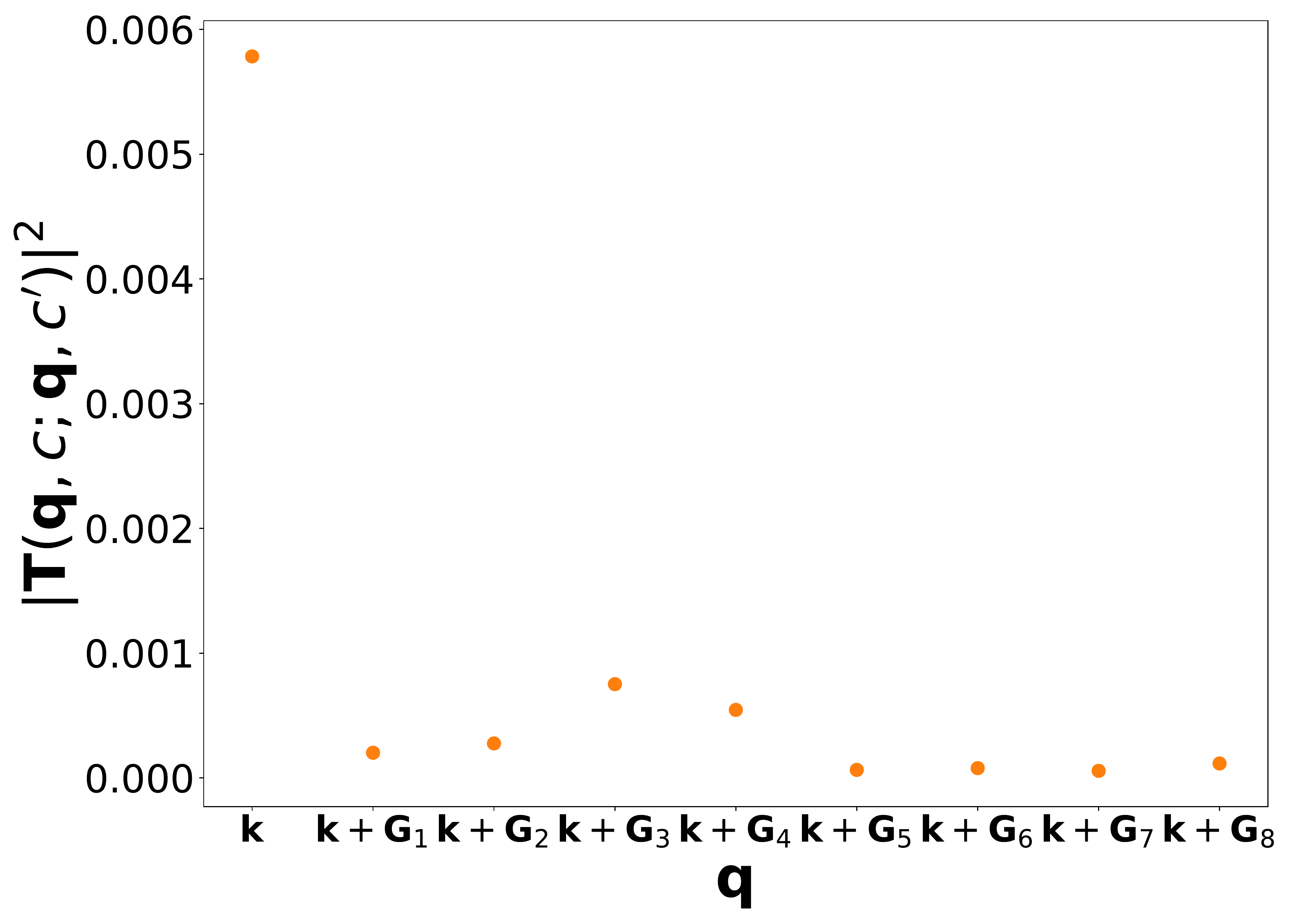}
\end{minipage}
\caption{\small{Square modulus of the overlap (defined in eq. (\ref{eq:Tknqm})) as a function of momentum $\bb{q}$ 
for a DFT-HSE calculation of hydrogen in the C2/c-24 structure at $T=200K$ and $P=250 GPa$. $\bb{G}_i$ are the reciprocal lattice vectors of the supercell, 
$\bb{k}$ corresponds to the $\bb{X+}$ vector of the primitive BZ (see Fig. \ref{fig:BST})
}\label{fig:Tknqm}}
\end{figure}

A non-vanishing $|\bb{T}(\bb{q},m;\bb{q},m')|^2$
indicates that the Born-Oppenheimer electronic wave function, $\Phi_\alpha^\Rvec$, 
averaged over the nuclear distribution transforms as an electronic wave function with
crystal momentum $\bb{q}$. 
Corrections beyond the harmonic approximation, as well as incomplete averaging over 
nuclear configurations may modify the results, but the matrix elements remain 
well peaked for a single crystal momentum,
see Fig.~\ref{fig:Tknqm} as an example.

For all the systems considered here, we have also verified that analysing
 $\bb{T}(\bb{q},m;\bb{q},m')$ using DFT-HSE wave function gives
the same quasi-momentum to the excitations as one would guess based on the
corresponding ideal crystal states $\Phi_{\bb{k}n}^{\bb{R}_0}(\bb{r})$. 

\subsection{\label{sec:Hydrogen}Results: Hydrogen}

In this section we report the results of the electronic structure for solid hydrogen in the C2/c-24 structure at 200 K and 248 GPa. 
The ideal crystalline structure information has been obtained by {\it ab initio} Random Structural Search method within the PBE approximation\cite{Pickard2007} 
and further relaxed at constant pressure using the DFT-vdW-DF functional. 
We include zero point motion of the protons using path integrals with the CEIMC algorithm at constant volume and temperature. 
We consider a nearly cubic supercell of 96 protons with $L_x=11.12 a_0, L_y=9.88 a_0, L_z=9.61 a_0$ where $a_0$ is the Bohr radius. 
Optimized Slater-Jastrow-backflow trial wave functions have been used for the CEIMC calculations \cite{Pierleoni2016}; details of the CEIMC simulations are reported in Ref.\cite{Rillo2018}. 
Averages over ionic positions for the electron addition and removal energies are obtained with 40 statistically independent configurations from the CEIMC trajectories.

For a given fixed nuclear configuration, the electron addition and removal energies are obtained by considering systems with a variable number of electrons $n=N_e-N_p \in [-6,6]$. 
For each system we perform Reptation Quantum Monte Carlo (RQMC) calculations with imaginary-time projection $t=$2.00 Ha$^{-1}$ 
and time step $\tau=$0.01 Ha$^{-1}$ for a $6\times 6\times 6$ Monkhorst-Pack grid of twists. 
We checked that those values are adequate for converging the addition/removal energies within our resolution. 
The electron addition and removal energies are further corrected for finite size effects in leading and next-to-leading order \cite{Yang2020}. 
\begin{figure*}[t]
\center
\begin{minipage}[b]{1.2\columnwidth}
\includegraphics[width=\columnwidth]{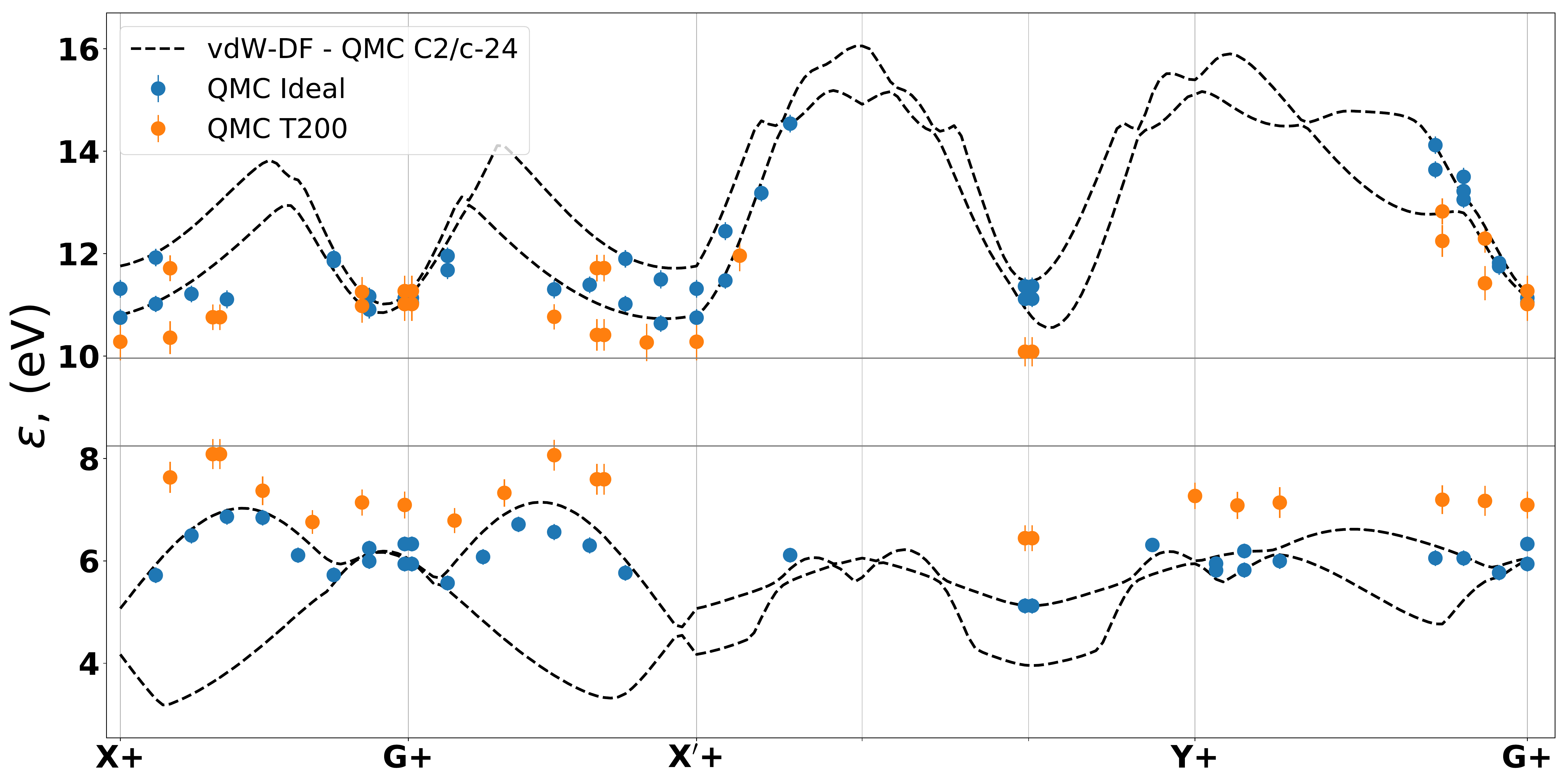}
\subcaption{\label{fig:BSTa}}
\end{minipage}
\begin{minipage}[b]{0.8\columnwidth}
\includegraphics[width=\columnwidth]{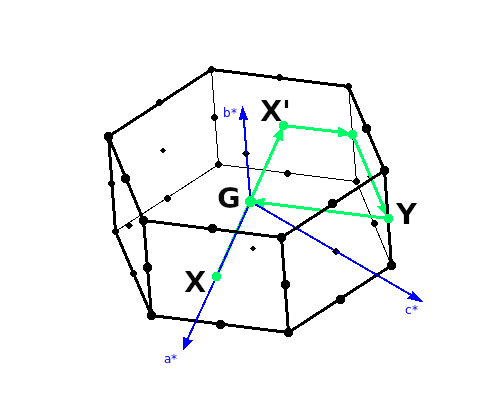}
\subcaption{\label{fig:BSTb}}
\end{minipage}
\caption{\small{(a) Band structure at finite temperature from QMC-CEIMC  (orange points with $6\times6\times6$ twist grid) 
and for perfect crystal QMC (blue points with $8\times8\times8$ twist grid) calculations of 96 hydrogen atoms compared with the band structure from vdW-DF density functional for a unit cell of C2/c-24 hydrogen crystal at 248 GPa.
The horizontal lines are the corresponding valence band maximum ($\mu^-$) and conduction band minimum ($\mu^+$) of the thermal crystal determined in Fig. \ref{fig:Nmu}. (b) The symmetry points of the Brillouin-Zone path. 
The actual path reported in panel (a) ($\bb{X+}, \bb{G+}, \bb{X'+}, \bb{Y+}$) 
was rigidly shifted  from the one indicated in panel (b) 
by ($\frac{2 \pi}{16L_x}$,$\frac{2 \pi}{16L_y}$,$\frac{2 \pi}{16L_z}$)
to better match the QMC twist grid.  
}\label{fig:BST}}
\end{figure*}

\begin{figure}
\center
\begin{minipage}[b]{\columnwidth}
\includegraphics[width=\columnwidth]{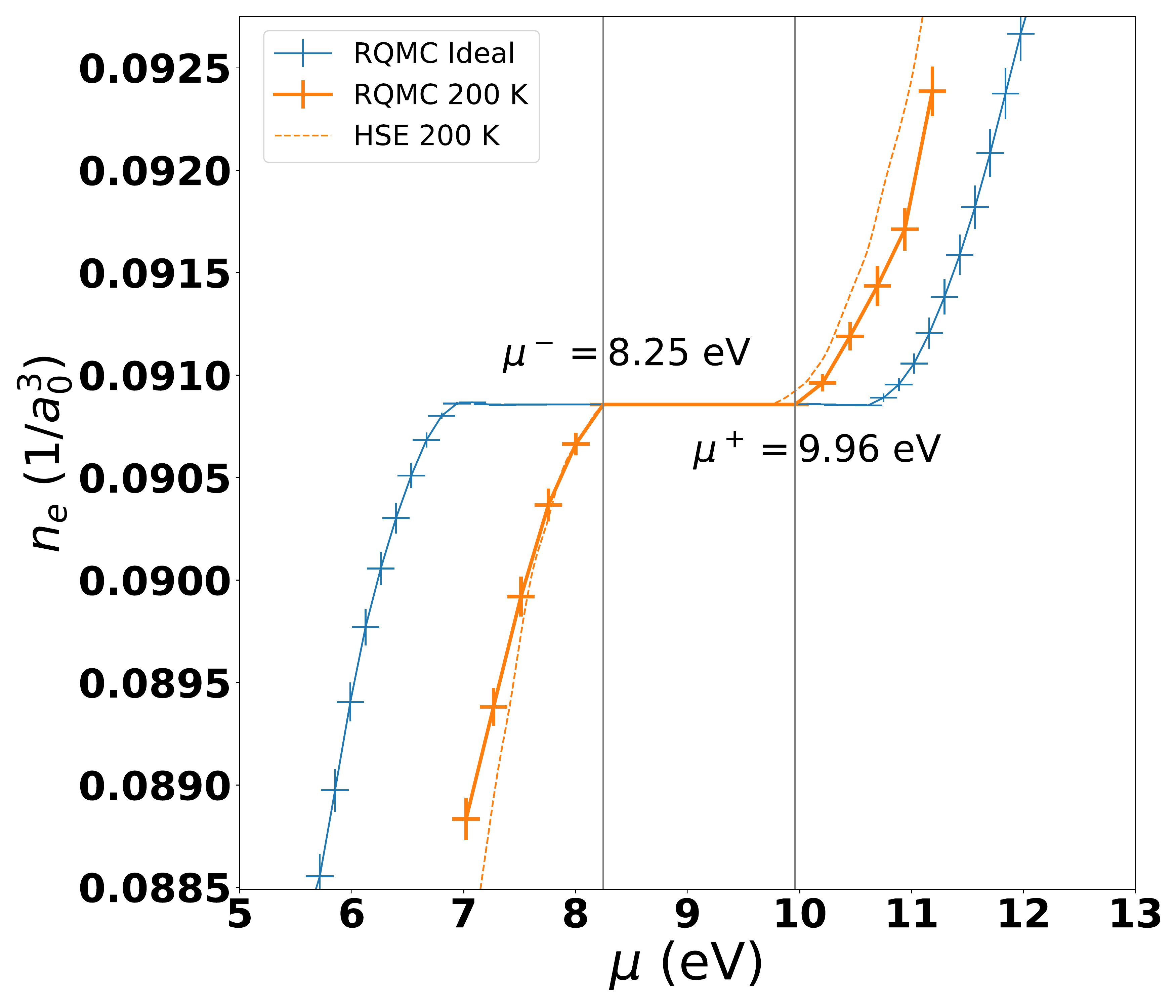}
\end{minipage}
\caption{\small{Mean electron density from QMC-CEIMC calculations (orange solid line) and the integrated DOS computed with HSE density functional (orange dashed line) 
for C2/c-24 hydrogen crystal at 248 GPa and 200 K plotted together with the RQMC  electron density for a perfect hydrogen crystal (blue line).
}\label{fig:Nmu}}
\end{figure}

Figure \ref{fig:BSTa} illustrates the QMC highest occupied and lowest unoccupied bands at 200 K and for the perfect crystal plotted 
on top of the DFT/vdW-DF band structure of the unitary cell of C2/c-24 hydrogen crystal at 250 GPa. The DFT band structure was rigidly shifted (``scissor operator'') to match the QMC gap. 
The QMC points are sparsely mapped onto the band structure of the primitive cell of crystalline hydrogen as we have only computed energies for a few twist values in order to determined the minimum insertion and removal energies 
for each twist value of our discrete grid of the supercell. 
Values for other momenta could be found by considering excitations at other twist values.

Figure \ref{fig:Nmu} shows the GCTABC results for the electron volume density $n_e$ defined in Eq.~(\ref{eq:n_e}) for solid hydrogen (with and without zero point motion).
We also report the DFT-HSE integrated DOS computed with the procedure described in section \ref{sec:GapGrandCan}. 
Vertical lines indicate the plateau region or, equivalently, the valence band maximum ($\mu^-$) and conduction band minimum ($\mu^+$) of the quantum crystal reported in Fig. \ref{fig:BSTa}. 

The reduction of the gap due to nuclear motion is of the order of 2 eV; this big change is caused by its large zero point motion. 
More details on the electronic gaps of thermal hydrogen crystals can be found in Ref.~\cite{Gorelov2020}. 

\section{\label{sec:OptProp}Optical properties at finite temperature}

In the previous section, we have discussed how to determine the fundamental gap and the
electronic band structure with explicitly correlated, many-body approaches, 
in particular with QMC methods. It is then natural to interpret the electron addition/removal
energies as electron/hole excitations with well defined (electronic) quasi-momentum in
the first BZ. The description of these excitations can be used in approximate single-particle theories, e.g. describing
linear response spectral functions. 
In the following, we will focus on the calculation of optical properties computed within the
Kubo-Greenwood (KG) formalism \cite{Kubo1957,Greenwood1958} taking into account the Born-Oppenheimer dynamics of the nuclei.

\subsection{Semi-classical averaging}

Let us assume that the exact electron-nuclear wave function can be factorized into
$ \ket{\alpha n} \simeq \ket{\Phi^{\mathbf{R}}_{\alpha}}\ket{\chi_{n}}$, where $\{\Phi^{\mathbf{R}}_{\alpha},E^{\mathbf{R}}_{\alpha}\}$ 
is the (Born-Oppenheimer) solution of the electronic problem that
depends parametrically on the nuclear configuration $\mathbf{R}$. 
Here $\{\ket{\chi_{n}},\Omega_n\}$ are the nuclear wave-functions
and energies, and $\epsilon^{\mathbf{R}}_{\alpha}=E^{\mathbf{R}}_{\alpha}-E^{\mathbf{R}}_{0}$ denotes the Born-Oppenheimer 
excitation energy. 
Let us assume that the nuclear eigenstates are well described by
the ground state Born-Oppenheimer energy surface $E_0(\mathbf{R})$, and neglect any
the dependence on electronic excitations $\alpha \ne 0$.

Since the electrons can be considered to be in the ground state for the temperatures considered here,
we can write the Kubo-Greenwood (KG) conductivity 
as a thermal average over nuclear states only,
\bea
    \sigma(\omega,T) = \frac{1}{Z} \sum_{n}  e^{-\frac{\Omega_{n}}{k_B T}} \sigma_n(\omega),
\label{eq:1condT}
\eea
where $Z=\sum_n e^{-\Omega_{n}/{k_B T}}$.
Within this  theory,
$\sigma_n(\omega)$ takes the form 
\bea
\sigma_n(\omega)&\propto&\frac{1}{\omega}\sum_{\alpha}^{occ.} \sum_{\beta,m}^{unocc.} 
|\bra{\chi_{n}} P^{\mathbf{R}}_{\alpha \beta} \ket{\chi_{m}}|^2 \nonumber \\
&&\times \delta(\epsilon_{\beta m}-\epsilon_{\alpha n} - \hbar \omega),
\label{eq:1exact}
\eea
where $\alpha$ indicates the initial electronic  states in the valence band $|\Phi^{\mathbf{R}}_{\alpha}\rangle$, 
 $\beta$ and $m$ are the final electronic and nuclear states in the conduction band, respectively,
 $P^{\mathbf{R}}_{\alpha \beta} = \bra{\Phi^{\mathbf{R}}_{\alpha}} \nabla \ket{\Phi^{\mathbf{R}}_{\beta}}$ 
the matrix element of the single electron momentum operator 
for a given  (fixed) nuclear configuration $\mathbf{R}$, 
and $\epsilon_{\alpha n} = \langle \chi_{n} | \epsilon^{\bb{R}}_{\alpha} | \chi_{n} \rangle+\Omega_n$ 
is the total electron-nuclear excitation energies.  

The conventional quasiclassical procedure introduced by Williams \cite{Williams1951} and Lax \cite{Lax1952} (WL) substitutes the nuclear states $m$ with a continuum
and replaces the eigenvalues $\epsilon_{\alpha n}$ in Eq. (\ref{eq:1exact}) by the 
electronic Born-Oppenheimer excitations for a fixed nuclear configuration $\epsilon^{\mathbf{R}}_{\alpha}$  (see refs. \cite{Lax1952,Patrick2014}) 
\bea
    \sigma^{WL}_n(\omega) \propto
\frac{1}{\omega}\sum_{\alpha}^{occ.} \sum_{\beta}^{unocc.} \bra{\chi_{n}} |P^{\mathbf{R}}_{\alpha \beta}|^2 \delta(\Delta \epsilon^{\mathbf{R}}_{\beta,\alpha} - \hbar \omega)  \ket{\chi_{n}} \nonumber \\
    \label{eq:SC}
\eea
with $\Delta \epsilon^{\mathbf{R}}_{\beta,\alpha} = \epsilon_{\beta}^\Rvec -\epsilon_{\alpha}^\Rvec$.
Using second order perturbation theory, it has been argued that this expression considers, in an effective way, the phonon-assisted indirect transitions\cite{Patrick2014,Zacharias2016}. 
However, for light nuclei such as hydrogen, replacing the nuclear spectrum by a classical continuum might not be accurate enough in the temperature regime explored by experiments.

\subsection{\label{sec:Quantum}Quantum averaging}

An alternative procedure is to consider only direct transitions between pairs of electronic states of 
quantum or thermally averaged bands taking into account only
nuclear zero point motion and thermal renormalizations of the bands, but
neglecting phonon assisted transitions. In practice, we replace the eigenvalues in Eq.(\ref{eq:1exact}) by their quantum/thermal average neglecting explicit phonon contributions.


Considering low temperatures, nuclear states will occupy
essentially only the  nuclear ground state and a few low-lying phonon excitations,
so that phononic energies can be neglected compared to typical electronic excitation energies
\beq
\epsilon_{\beta m} - \epsilon_{\alpha n} \approx \epsilon_{\beta n}-\epsilon_{\alpha n}
\eeq
The summation over nuclear states $m$ can be replaced using the completeness relation $\sum_m |\chi_{\beta m}\rangle\langle\chi_{\beta m}|=1$,
\bea
\sigma_n(\omega) &\propto& \frac{1}{\omega} \sum_{\alpha}^{occ.} \sum_{\beta}^{unocc.}  \bra{\chi_{\alpha n}} |P^{\mathbf{R}}_{\alpha \beta}|^2 \ket{\chi_{\alpha n}} \nonumber \\ && \times  \delta(\epsilon_{\beta n}-\epsilon_{\alpha n}-\hbar \omega). 
\label{eq:1approx1}
\eea
Within CEIMC the nuclear configurations, including zero point motion and phononic excitation at finite temperature, are sampled exactly
according to the Born-Oppenheimer PES without any explicit phonon sampling, so that Eq.~(\ref{eq:1approx1}) cannot be applied directly. 
However, at zero temperatures, the nuclear average involved in calculating the matrix elements
fully separates from the nuclear average in the electronic excitation energies. Assuming this
factorization to hold also at low, but finite temperatures, we obtain
\bea
\label{eq:QA}
\sigma(\omega,T) &\propto& \frac{1}{\omega } \sum_{\alpha}^{occ.} \sum_{\beta}^{unocc.}  \langle |P^{\mathbf{R}}_{\alpha \beta}|^2 \rangle 
 \nonumber \\ && \times \delta(\langle\Delta \epsilon_{\alpha \beta}\rangle-\hbar \omega)  \nonumber \\  
\eea
We have obtained effectively a Kubo-Greenwood conductivity, where the $\delta$ function is represented as a gaussian with the eigenvalues and matrix elements averaged over the nuclear states.
 Eq. (\ref{eq:QA}) is valid at low temperatures and when the transition matrix elements do not correlate with the eigenvalues. 
This effectively means that the transitions are computed 
for electronic eigenvalues averaged over the nuclear motion. We will call this procedure Quantum Averaging. 

\subsection{\label{sec:HydrCarb}Results: crystalline hydrogen and carbon diamond}

\begin{figure*}[t]
\center
\begin{minipage}[b]{\columnwidth}
\includegraphics[width=\columnwidth]{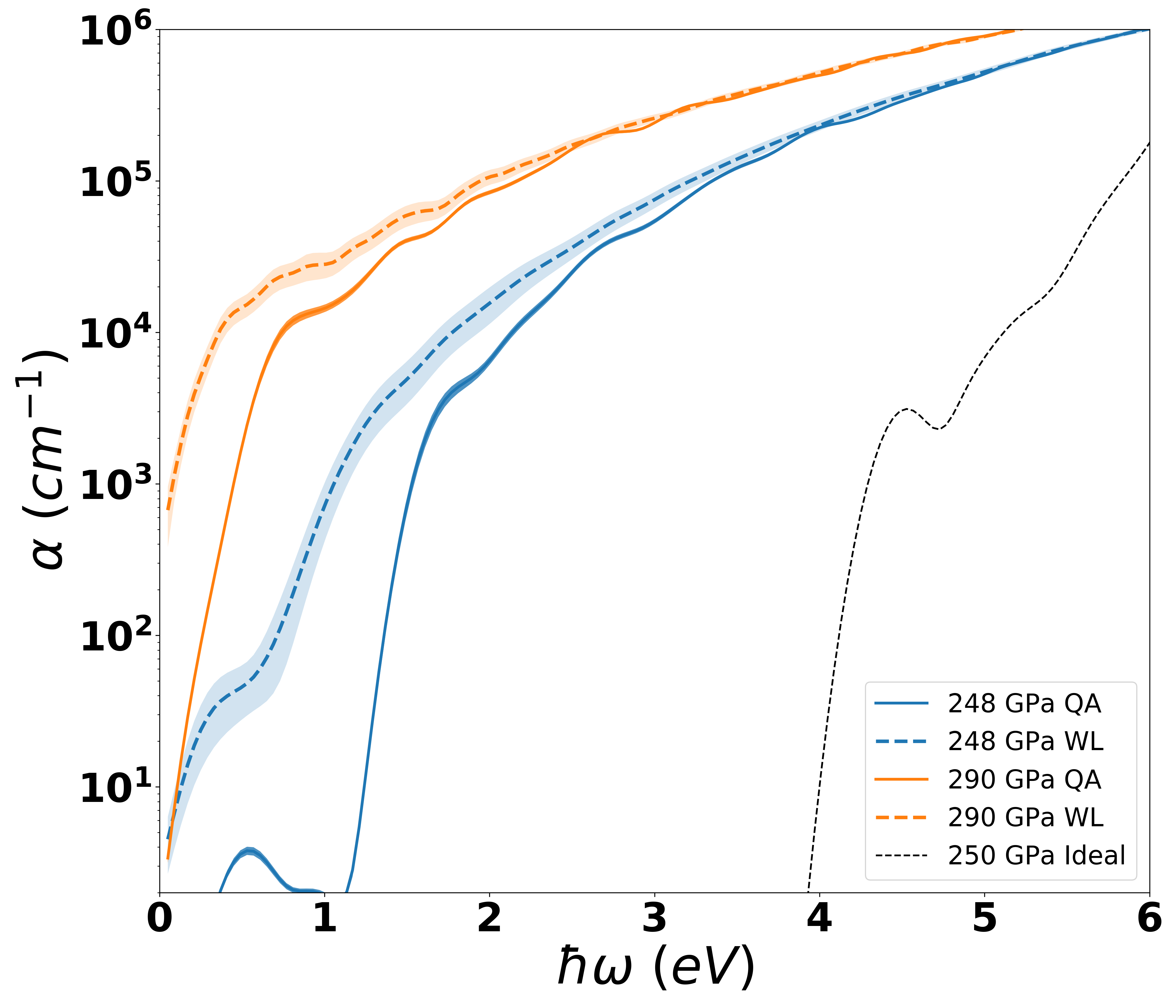}
\subcaption{Hydrogen \label{fig:QASCa}}
\end{minipage}
\begin{minipage}[b]{\columnwidth}
\includegraphics[width=\columnwidth]{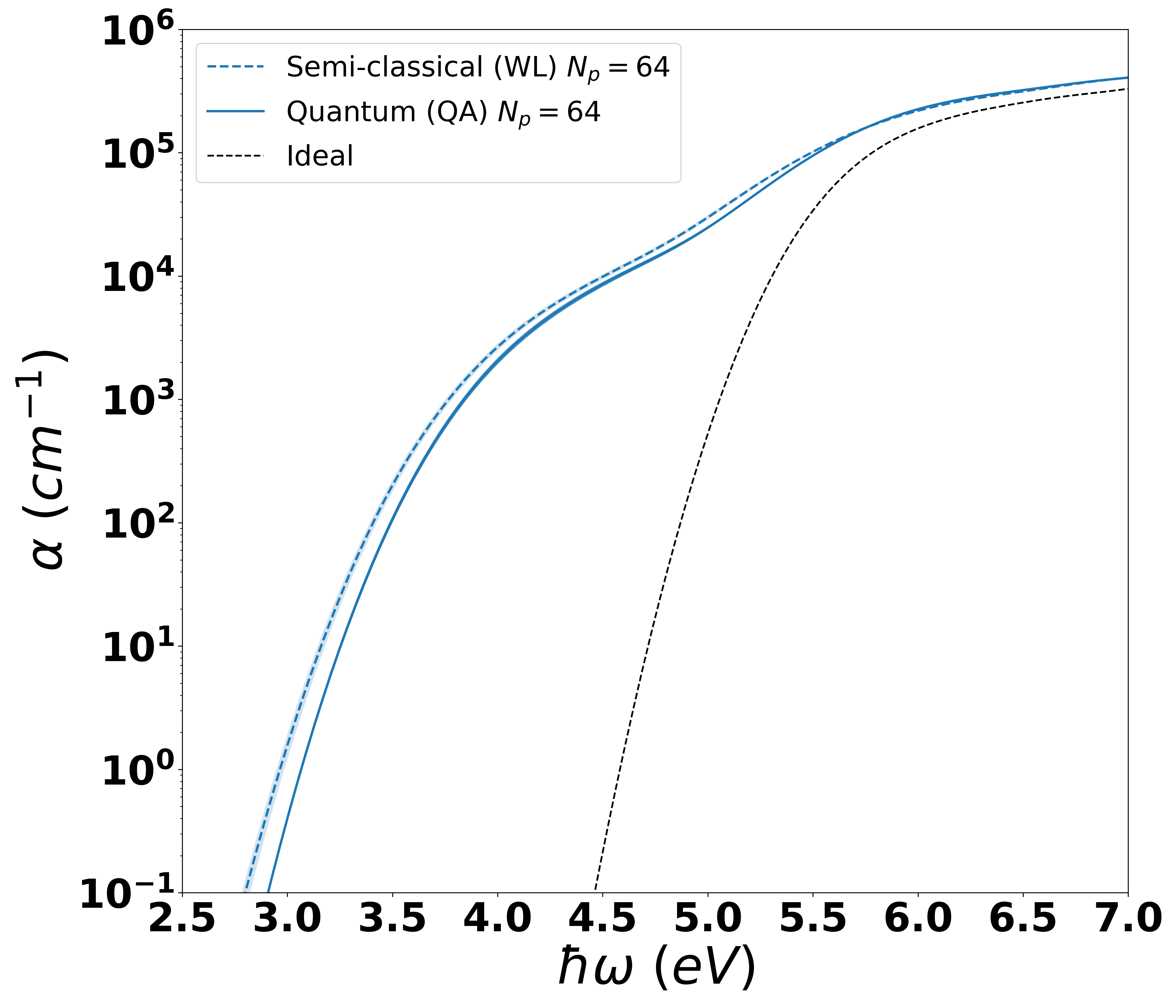}
\subcaption{Carbon \label{fig:QASCb}}
\end{minipage}
\caption{\small{ Optical absorption for (a) C2/c-24 quantum crystals at 200K and (b) carbon diamond at 297 K using the semiclassical (WL) and quantum (QA) averaging procedures. 
The black dashed lines indicate the absorption of an ideal crystal. 
}\label{fig:QASC}}
\end{figure*}

Here we illustrate the different renormalization procedures of the absorption spectra for solid C2/c-24 hydrogen at 200 K and for carbon diamond at 297 K. 
While nuclear configurations of hydrogen are obtained as described in section \ref{sec:Hydrogen} 
for carbon diamond we used the cubic supercell containing 64 atoms with the lattice constant 3.56712$\angstrom$ appropriate for room temperature \cite{Stoupin2010}. 
The nuclear configurations for carbon are obtained from the path integral molecular dynamics with the PBE functional using i-PI and QuantumEspresso \cite{iPI,QE2017}. 

The absorption is computed for 40 fixed nuclear configuration within the Kubo-Greenwood formalism implemented in KGEC code \cite{Calderin2017a} using DFT. 
In hydrogen we used HSE functional with $8\times8\times8$ k-grid and $2\times2\times2$ q-grid to sample the Fock operator, 
the kinetic energy cut-off was set to 40 Ry and the gaussian smearing to 0.2 eV. 
In Carbon we used PBE functional with $10\times10\times10$ k-grid, 60 Ry of the cut-off and 0.35 eV smearing. For both systems the PAW pseudopotentials were used. 

Figure \ref{fig:QASCa} shows the absorption spectra of hydrogen at 248 and 290 GPa together with the absorption of ideal crystal at 250 GPa, 
computed  with the HSE functional at the same conditions as for quantum crystals. The difference between the semiclassical and quantum averaging is particularly noticeable in the region of the onset of absorption (low energy/low absorption). 
The reduction of the gap at 248 GPa, given by the difference in the onset of the ideal and quantum absorption profiles is 
compatible with the reduction obtained by QMC in section \ref{sec:Hydrogen}.  

\begin{figure*}[t]
\center
\begin{minipage}[b]{\columnwidth}
\includegraphics[width=\columnwidth]{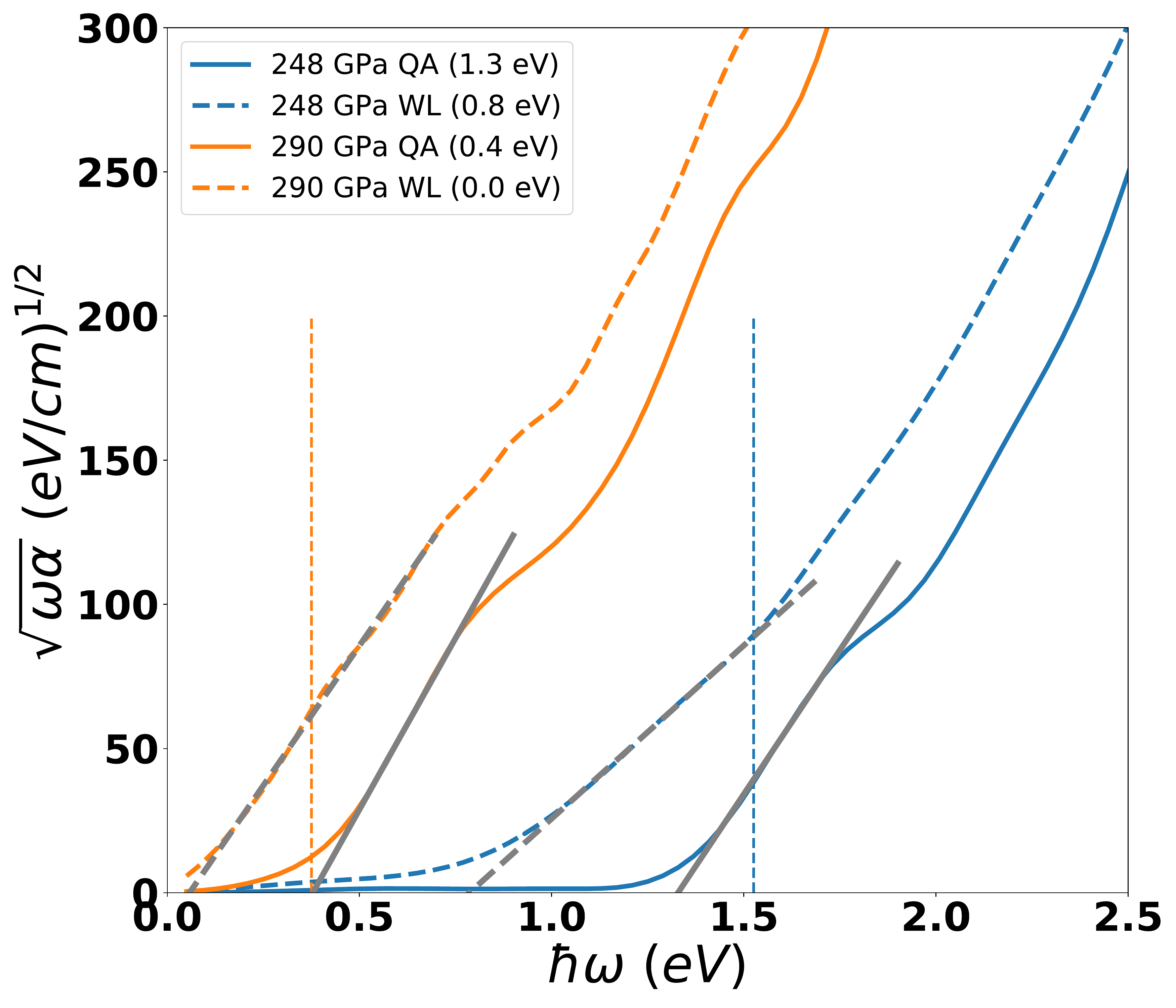}
\subcaption{Hydrogen \label{fig:Tauca}}
\end{minipage}
\begin{minipage}[b]{\columnwidth}
\includegraphics[width=\columnwidth]{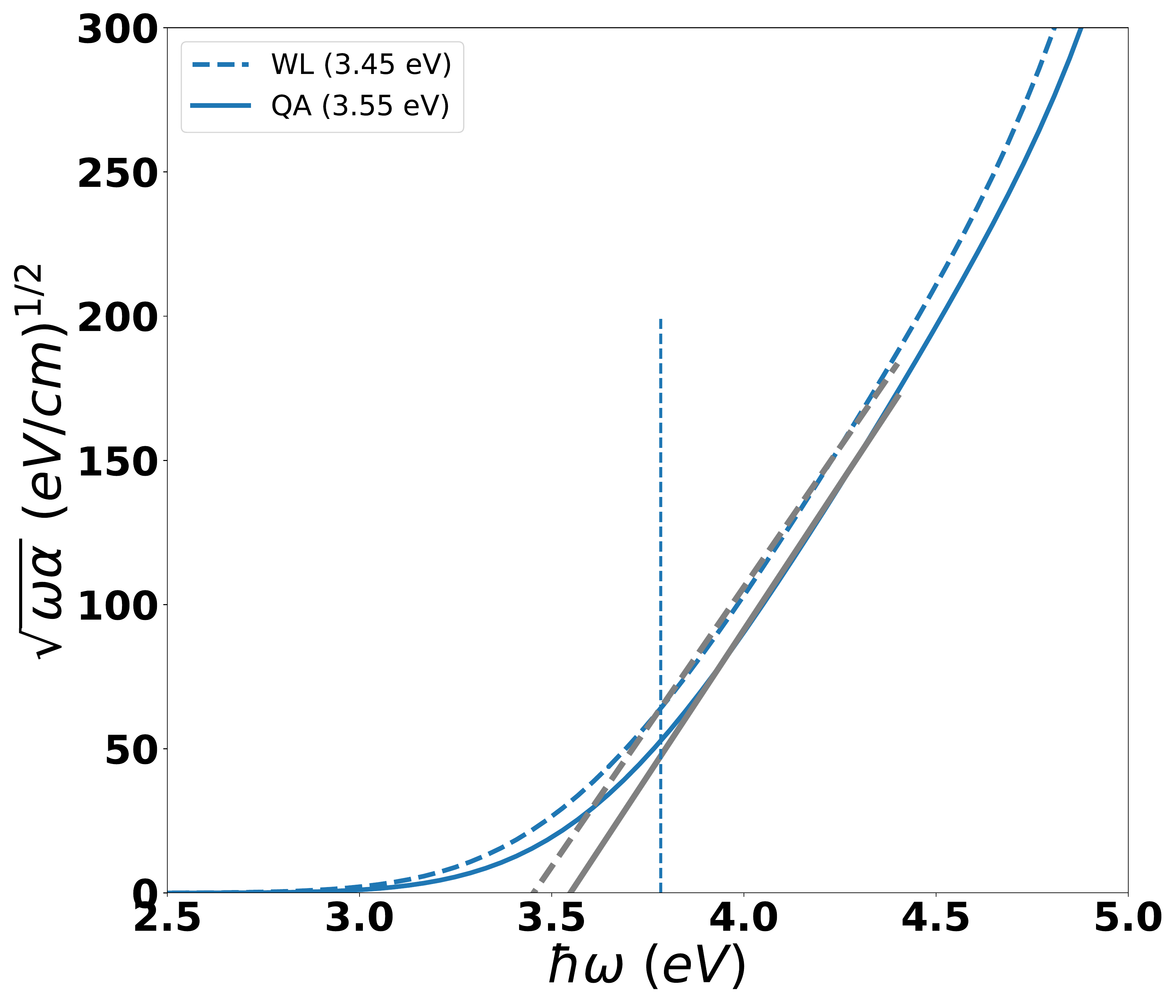}
\subcaption{Carbon \label{fig:Taucb}}
\end{minipage}
\caption{\small{Tauc analysis of the absorption profiles for (a) C2/c-24 quantum crystals at 200K and (b) carbon diamond at 297 K using the semiclassical (WL) and quantum (QA) averaging procedures. 
Values of the gap extracted from the intercept of the linear fits and the
horizontal axis are reported in the legend. The dashed vertical lines indicate the values of fundamental band-gaps computed using the HSE functional for hydrogen (a) and PBE for carbon (b). 
}\label{fig:Tauc}}
\end{figure*}

The fundamental gap is often determined in experiments by using the Tauc extrapolation of absorption profiles \cite{Tauc1966}. 
Figure \ref{fig:Tauca} illustrates the Tauc analysis of our absorption profiles for C2/c-24 hydrogen at 200K at two pressures. 
The gap value is determined from the value of the intercept of a linear fit of $\sqrt{\omega \alpha}$ vs $\omega$ with the horizontal axis. At 290 GPa, 
the semi-classical procedure predicts the gap to be closed
at variance with the fundamental gap of $\sim0.4$ eV from the HSE band structure (orange vertical dashed line). 
The absorption profile by Quantum Averaging provides a more consistent value of the gap. At 248 GPa, the gap from the semi-classical absorption is $\sim 0.8$ eV, 
about $0.5$ eV smaller than the fundamental gap from the HSE band structure (blue vertical dashed line). 
Again the gap extracted from the absorption profile by quantum averaging is in better agreement with the one from the band structure. 

The results for carbon, which are shown in Fig.~\ref{fig:QASCb}, demonstrate a smaller reduction from ideal to thermal crystal absorption. 
The results are compatible with the ones by Zachrias et al. 
Ref.\cite{Zacharias2016} up to a horizontal shift in the absorption profile used there to match the GW band gap. 
Since carbon is heavier than hydrogen, one expect the semiclassical averaging to become more accurate. From Fig. \ref{fig:QASCb} we see that 
the difference between semiclassical and quantum averaging profiles is much smaller than in hydrogen. The gaps extracted from the absorption profiles for carbon, see Fig. \ref{fig:Taucb}, are also compatible with the gap extracted from the PBE band structure.
Note, however, that the gap values extracted from the Tauc analysis depend on the gaussian smearing in computing the absorption profile and on the fitting range used.

\section{\label{sec:Concl}Conclusions}

In this work we have discussed electronic structure and optical properties for quantum crystals where quantum and thermal nuclear motion is important. We have focused on low temperatures assuming the validity of the Born-Oppenheimer approximation.
First, we addressed the fundamental electronic gap in terms of electron addition and removal energies which can be directly computed using many-body
Quantum Monte Carlo methods without relying on perturbation theory. We presented the methodology in the canonical and grand-canonical ensembles, the latter giving access to density of states. 
The averaging over motion of the ions is based on their thermodynamic definitions, e.g. the electronic (energy) density $n_e(\mu)$ ($e(\mu)$). 
Therefore, the resulting energy gap
might be different from a semi-classical approach where addition and removal energies
for individual nuclear configurations are considered independently.

We then discussed how the electronic band-structure can be established for quantum crystals assigning to each addition/removal energy a well-defined crystal momentum  for averages over nuclear equilibrium distributions. 
We presented an explicit procedure  to determine the
crystal momentum for electronic wave functions in the Born-Oppenheimer approximation without
the need to explicitly fix the phase of the BO nuclear wavefunction.

We then discussed the calculation of optical conductivity using the Kubo-Greenwood formula. 
For light nuclei at low temperatures, we have shown that the semi-classical approximation should be replaced by a ''quantum average'' 
which takes into account the quantization of the nuclear motion, similar to the fundamental gap discussed before.

We have applied our methodology to study the band structure and density of states of hydrogen C2/c-24 crystal at 200K and 250GPa where CEIMC has been used to sample the nuclear density
matrix, and the QMC methods for the addition/removal energies.

Concerning optical properties, we have calculated the optical absorption for hydrogen and carbon diamond where we have approximated the electronic Born-Oppenheimer energies by DFT Kohn-Sham eigenvalues 
and discussed the differences between the quantum average and the semi-classical approximation.
At low temperatures, only the quantum averaging procedure correctly reproduces the onset of absorption
consistent with the one expected from the fundamental energy gap.
As expected, in the case of carbon, the difference between the two methods are much smaller than for hydrogen.

This work paves the way for a more controlled and thermodynamically consistent investigation of electronic structure from many-body theory.
Such procedures are particularly needed for strongly correlated systems which have important thermal and quantum ionic motions.  
The optical responses of such materials are one of the most basic and accurate experimental probes of their electronic properties. Hence simulation methods need to be able to consider all important effects to make unambiguous comparisons with experiment. 

\begin{acknowledgments}
This work has received funding from the European Union’s Horizon  2020  research  and  innovation  program  under  the grant agreement No.  676531 (project E-CAM). D.M.C. was supported by DOE Grant NA DE-NA0001789 and by the Fondation NanoSciences (Grenoble). V.G. and C.P. were supported by the Agence Nationale de la Recherche (ANR) France, under the program ``Accueil de Chercheurs de Haut Niveau 2015'' project: HyLightExtreme. 
Computer time was provided by the PRACE Project 2016143296, ISCRAB (IsB17\_MMCRHY,IscrB\_FPSHPH) computer allocation at CINECA Italy, the high-performance computer resources from Grand Equipement National de Calcul Intensif (GENCI) Allocation 2018-A0030910282, and by
the Froggy platform of CIMENT, Grenoble (Rh{\^o}ne-Alpes CPER07-13 CIRA and ANR-10-EQPX-29-01).
\end{acknowledgments}

\section*{Data Availability}
The data that support the findings of this study are available from the corresponding author upon reasonable request.

\appendix 

\section{Grand-canonical twist averaging boundary conditions}

Here we give the procedure to apply the grand-canonical twist averaging boundary conditions (GCTABC) \cite{Chiesa2006} 
for the total energy calculations with varying numbers of electrons and twist angles. At each twist angle, $\theta$, the electronic wave function
obeys \cite{Lin2001} 
\bea
\Psi(\bb{r}_1+L_x \hat{x}..,\bb{r}_{N_e})=e^{i\theta_x}\Psi(\bb{r}_1,..,\bb{r}_{N_e}).
\eea
and is used to calculate the $N_e$ electron ground state energy $E_0(N_e,\theta)$.

At zero (electronic) temperature in the grand-canonical ensemble, the probability for a number, $N^{\theta}_e)$ of electrons for a given twist angle $\theta$ is proportional to
\bea
\label{rhs}
\mathcal{P}(N^{\theta}_e) &\propto& \exp [-\beta (E_0(N_e,\theta)-\mu N_e)] \\ \nonumber &\xrightarrow[\beta \to \infty]{}& \min_{N_e} [E_0(N_e,\theta)-\mu N_e],
\eea
where $\mu$ is the chemical potential. 

In practice, within GCTABC, the ground state energy $E_0(N_e,\theta)$ is computed for different number of electrons, $N_e$, 
independently of the twist vector $\theta$. A uniform background charge is used to ensure global charge neutrality. 
Given $\mu$ and $\theta$ we can then determine $N^{\theta}_e$ from Eq.~(\ref{rhs}).
Equivalently, since $\partial E_0(N_e,\theta)/\partial N_e=\mu$ at the minimum of Eq.~(\ref{rhs}),
we can also scan the energy differences for each value of the chemical potential, $\mu$, and determine $N^{\theta}_e$ from bracketing
\bea
E_0(N_e,\theta)-E_0(N_e-1,\theta) \leqslant \mu < E_0(N_e+1,\theta)-E_0(N_e,\theta), \nonumber \\
\eea
resulting in an optimal number of electrons at each value of the chemical potential and twist $N^{\theta}_e(\mu)$. 

To reduce finite size effects we further average over $M$ twist angles and divide by the volume of the supercell to obtain
\bea
e(\mu)&=&\frac{1}{MV}\sum_{\theta}E_0(N^{\theta}_e,\theta) \\
n_e(\mu)&=&\frac{1}{MV}\sum_{\theta}N^{\theta}_e(\mu)
\eea 
the electronic and energy density $n_e(\mu)$ and $e_0(\mu)$. 

The resolution of the densities depends on the underlying twist grid, but the resolution of the
chemical potential depends on the statistical error.

The twist grid error on the energy can be  improved by calculating the Legendre transform of the grand potential as described in Ref.~\cite{PhysRevB.100.245142} or may be
fully eliminated by using symmetries and twist pockets as described in refs. \cite{Chiesa2006,Holzmann2016}.

\bibliography{aipsamp}

\end{document}